\documentstyle[12pt]{article}
\def\hybrid{\topmargin -20pt	\oddsidemargin 0pt
	\headheight 0pt	\headsep 0pt
	\textwidth 6.25in	% A4 paper
	\textheight 9.5in	% A4 paper
	\marginparwidth .875in
	\parskip 5pt plus 1pt	\jot = 1.5ex}
\def\baselinestretch{1.2}

\catcode`\@=11

\def\marginnote#1{}
\newcount\hour
\newcount\minute
\newtoks\amorpm
\hour=\time\divide\hour by60
\minute=\time{\multiply\hour by60 \global\advance\minute by-\hour}
\edef\standardtime{{\ifnum\hour<12 \global\amorpm={am}%
	\else\global\amorpm={pm}\advance\hour by-12 \fi
	\ifnum\hour=0 \hour=12 \fi
	\number\hour:\ifnum\minute<10 0\fi\number\minute\the\amorpm}}
\edef\militarytime{\number\hour:\ifnum\minute<10 0\fi\number\minute}
\def\draftlabel#1{{\@bsphack\if@filesw {\let\thepage\relax
   \xdef\@gtempa{\write\@auxout{\string
      \newlabel{#1}{{\@currentlabel}{\thepage}}}}}\@gtempa
   \if@nobreak \ifvmode\nobreak\fi\fi\fi\@esphack}
	\gdef\@eqnlabel{#1}}
\def\@eqnlabel{}
\def\@vacuum{}
\def\draftmarginnote#1{\marginpar{\raggedright\scriptsize\tt#1}}

\def\draft{\oddsidemargin -.2truein
	\def\@oddfoot{\sl preliminary draft \hfil
	\rm\thepage\hfil\sl\today\quad\militarytime}
	\let\@evenfoot\@oddfoot	\overfullrule 3pt
	\let\label=\draftlabel
	\let\marginnote=\draftmarginnote
   \def\@eqnnum{(\theequation)\rlap{\kern\marginparsep\tt\@eqnlabel}%
\global\let\@eqnlabel\@vacuum}  }

\def\preprint{\twocolumn\sloppy\flushbottom\parindent 2em
	\leftmargini 2em\leftmarginv .5em\leftmarginvi .5em
	\oddsidemargin -.5in	\evensidemargin -.5in
	\columnsep .4in	\footheight 0pt
	\textwidth 10.in	\topmargin  -.4in
	\headheight 12pt \topskip .4in
	\textheight 6.9in \footskip 0pt
	\def\@oddhead{\thepage\hfil\addtocounter{page}{1}\thepage}
	\let\@evenhead\@oddhead	\def\@oddfoot{}	\def\@evenfoot{} }

\def\numberbysection{\@addtoreset{equation}{section}
	\def\theequation{\thesection.\arabic{equation}}}

\def\underline#1{\relax\ifmmode\@@underline#1\else
	$\@@underline{\hbox{#1}}$\relax\fi}

\def\titlepage{\@restonecolfalse\if@twocolumn\@restonecoltrue\onecolumn
     \else \newpage \fi \thispagestyle{empty}\c@page\z@
	\def\thefootnote{\fnsymbol{footnote}} }

\def\endtitlepage{\if@restonecol\twocolumn \else \newpage \fi
	\def\thefootnote{\arabic{footnote}}
	\setcounter{footnote}{0}}  %\c@footnote\z@ }

\catcode`@=12
\relax

\def\figcap{\section*{Figure Captions\markboth
	{FIGURECAPTIONS}{FIGURECAPTIONS}}\list
	{Figure \arabic{enumi}:\hfill}{\settowidth\labelwidth{Figure
999:}
	\leftmargin\labelwidth
	\advance\leftmargin\labelsep\usecounter{enumi}}}
 \relax
\def\tablecap{\section*{Table Captions\markboth
	{TABLECAPTIONS}{TABLECAPTIONS}}\list
	{Table \arabic{enumi}:\hfill}{\settowidth\labelwidth{Table
999:}
	\leftmargin\labelwidth
	\advance\leftmargin\labelsep\usecounter{enumi}}}
 \relax
\def\reflist{\section*{References\markboth
	{REFLIST}{REFLIST}}\list
	{[\arabic{enumi}]\hfill}{\settowidth\labelwidth{[999]}
	\leftmargin\labelwidth
	\advance\leftmargin\labelsep\usecounter{enumi}}}
 \relax
\makeatletter
\newcounter{pubctr}
\def\publist{\@ifnextchar[{\@publist}{\@@publist}}
\def\@publist[#1]{\list
	{[\arabic{pubctr}]\hfill}{\settowidth\labelwidth{[999]}
	\leftmargin\labelwidth
	\advance\leftmargin\labelsep
	\@nmbrlisttrue\def\@listctr{pubctr}
	\setcounter{pubctr}{#1}\addtocounter{pubctr}{-1}}}
\def\@@publist{\list
	{[\arabic{pubctr}]\hfill}{\settowidth\labelwidth{[999]}
	\leftmargin\labelwidth
	\advance\leftmargin\labelsep
	\@nmbrlisttrue\def\@listctr{pubctr}}}
 \relax
\makeatother
\newskip\humongous \humongous=0pt plus 1000pt minus 1000pt

\newif\ifdtup

\relax
\hybrid
\def\thefootnote{\fnsymbol{footnote}}
\def\be{\begin{equation}}
\def\ee{\end{equation}}
\def\ba{\begin{eqnarray}}
\def\ea{\end{eqnarray}}

\begin{document}
\renewcommand{\theequation}{\thesection.\arabic{equation}}
\newcommand{\beq}{\begin{equation}}
\newcommand{\eeq}[1]{\label{#1}\end{equation}}
\newcommand{\ber}{\begin{eqnarray}}
\newcommand{\eer}[1]{\label{#1}\end{eqnarray}}
\begin{titlepage}
\begin{center}

\hfill CERN-TH.7332/94\\
\hfill LPTENS-94/19\\
\hfill hep-th/9407004\\

\vskip .5in

{\large \bf Gravitational Phenomena From Superstrings in Curved
Spacetime\footnote{Talk presented by C. Kounnas in the XXIX Moriond
Meeting, M\'eribel, France, 1994}}
\vskip .6in

{\bf Elias Kiritsis and Costas Kounnas\footnote{On leave from Ecole
Normale Sup\'erieure, 24 rue Lhomond, F-75231, Paris, Cedex 05,
FRANCE.}}\\
\vskip
 .4in

{\em Theory Division, CERN, CH-1211\\
Geneva 23, SWITZERLAND} \footnote{e-mail addresses:
KIRITSIS,KOUNNAS@NXTH04.CERN.CH}\\

\vskip .5in

\end{center}

\begin{center} {\bf ABSTRACT } \end{center}
\begin{quotation}\noindent
The four-dimensional superstring solutions define at  low energy
effective supergravity theories. A class of them extends successfully
 the validity of the standard model up to the string scale (${\cal
O}(10^{17})~TeV$). We stress the importance of string corrections
which are relevant for low energy (${\cal O}(1)~TeV$) predictions of
gauge and Yukawa couplings as well as the spectrum of the
supersymmetric particles.
A class of exact string solutions are also  presented, providing non
trivial space-time backgrounds, from which we can draw some lessons
concerning the regions of space-time where the notion of the
effective field theory prescription make sense. We show that the
string gravitational phenomena may induce during the cosmological
evolution, transitions from one effective field theory prescription
to a different one where the geometrical and topological data, as
well as the relevant observable states are drastically  different.

\end{quotation}
\vskip 1.0cm
CERN-TH.7332/94 \\
April 1994\\
\end{titlepage}
\vfill
\eject
\def\baselinestretch{1.2}
\baselineskip 16 pt
\noindent
\section{Introduction}
\setcounter{equation}{0}

String theory extents the validity of quantum field theory to very
short distances and defines consistently quantum gravity. It is thus
appropriate to try to investigate the behavior of string
dynamics at high energies and in regions of spacetime where the
gravitational field is strong.  There are several
problems in gravity where the classical, and even worse the
semiclassical treatment have perplexed physicists for decades. We are
referring here to questions concerning the behavior in regions of
strong (or infinite ) curvature with both astrophysical
(black holes) and cosmological (big-bang, wormholes) interest.
It is only appropriate to try to elucidate such questions in the
context of stringy gravity. There has been progress towards this
direction, and by now we have at least some ideas on how different
string gravity can be from general relativity in regions of
space-time with strong curvatures. We need however exact classical
solutions of string theory in order to have more quantitative control
on phenomena that are characteristic of stringy gravity.

On the other hand the special characteristics of superstrings do not
only reflect in the gravitational sector of the theory.
There are also important implications for particle physics, namely,
concerning the string low energy predictions at $M_Z$ and at the
accessible by the future, energy scale of ${\cal O}(1)$ $TeV$.

 The first main property of superstings is that they are ultraviolet
finite theories (at least perturbatively). The second important
property is that  they unify gravity with all other interactions.
This unification does not include  only the gauge interactions but
also the  Yukawa  ones as well as the interactions among the scalars.
This String Hyper-Unification (SHU) happens  at  large energy scales
$E_t={\cal O}(M_{string})=10^{17}~GeV$. At this energy scale however,
the first exited string states become important and thus, the whole
effective low energy field theory picture breaks down. Indeed, the
effective field theory of strings is valid only for  $E_t \ll
M_{string}$ by means of
${\cal O}(E_t/M_{string})^2$ expansion. It is then necessary to
evolve the SHU predictions to a lower scale $M_U < M_{string}$ where
the  effective field theory picture makes sense. Then, at $M_U$ any
string solution, provides  non-trivial relations among the gauge and
Yukawa couplings which can be written as,

\begin{equation}
\frac{k_i}{\alpha_i(M_U)}=\frac{k_j}{\alpha_j(M_U)}+\Delta_{ij}(M_U).
\label{shu} \end{equation}

The above relation looks very similar to the well known unification
condition
in supersymmetric Grand Unified Theories (SuSy-GUT) where the
unification scale is about $M_U=10^{16}~GeV$ and
$\Delta_{ij}(M_U)=0$ in the ${\bar {DR}}$ renormalization scheme; in
SuSy-GUTs the normalization constants $k_i$ are fixed $only$ for the
gauge couplings ($k_1=k_2=k_3=1$, $k_{em}=\frac{3}{8}$) but there are
no relations among gauge and Yukawa couplings at all. In string
effective theories however, the normalization constants ($k_i$) are
known for both gauge and Yukawa interactions. Furthermore,
$\Delta_{ij}(M_U)$ are calculable $finite$ quantities for any
particular string solution. Thus, the predictability of a given
string solution is extended for all low energy coupling constants
${\alpha_i(M_Z)}$ once the string induced corrections
$\Delta_{ij}(M_U)$ are determined.

 This determination  however, requests string computations which at
present are not completely known. It turns out that
$\Delta_{ij}(M_U)$ are non trivial functions of the vacuum
expectation values  of  some gauge singlet fields\cite{g1},
$~~<T_A>=t_A$, the so called moduli (the moduli fields  are flat
directions at the string classical level and they remain flat in
string perturbation theory, in the exact supersymmetric limit) :
\begin{equation}
{\Delta_{ij} (M_U)= {\delta_{ij}}+ F_{ij} (t_A)}.
\label{delta} \end{equation}
$F_{ij}(t_A)$ are modular forms  which depend on the particular
string solution and are normalized here such that:
$F_{ij}(t_A)=0$ when $t_A=1/M^2_{string}$.  Partial results for
$F_{ij}$ exist in the exact supersymmetric limit in many string
solutions based on orbifold
\cite{orb} and fermionic constructions \cite{ferm}.
 The finite part $\delta_{ij}$
is a function of $M_U/M_{string}$ and at the  present time it is only
 approximately estimated; in principle it is a well defined
calculable quantity once we perform our calculations  at the string
level where all interactions including gravity are  consistently
defined. The full string corrections   to the coupling constant
unification, $\Delta_{ij}(M_U)$, as well as the string corrections
associated to the soft supersymmetry  breaking parameters
 \begin{equation}
m_0,~~ m_{1/2},~~ A,~~ B~~{\rm  and}~~ \mu,~~~{\rm  at}~~~~M_U,
\label{soft} \end{equation}
are of main importance, since they fix  the strength of the gauge and
Yukawa interactions \cite{anto,fkz}, the full spectrum of the
supersymmetric particles as well as the Higgs and the top-quark
masses at the low energy range $E_t\sim M_Z$ - ${\cal O}(1)~TeV$.

There has been  a lot of progress in this direction and some
semi-quantitative results are already obtained. A much more detailed
study is necessary in order to understand better the SHU-predictions
when supersymmetry is spontaneously broken.

In the rest of this talk, however, we will focus mostly on some of
the implications of superstring theory for gravitational type
phenomena.
In particular we will analyze some exact four dimensional string
solutions which have been  constructed recently and they posses
interesting cosmological and astrophysical properties.

Studies on string theory have so far given hints for the presence of
several interesting stringy phenomena, like the existence of a
minimal
distance \cite{md}, finiteness at short distances \cite{m}, smooth
topology change \cite{tc,gk}, spacetime duality
symmetries \cite{r,bu,k1,GR,k2,AG,gk,kk11}, variable
dimensionality of spacetime \cite{di}, existence of maximal
(Hagedorn) temperature and subsequent phase transitions \cite{PT}
etc. The lessons we learn from the exact solutions we are going to
describe essentially corroborate some items on the list above and we
would like to present them in a somewhat general context. Ideas of a
similar form have already been presented in \cite{kk}.

In particular, we will present exact string solutions, in 3+1
dimensions, where one can study a topology and/or geometry changing
phenomena which can happen in some regions of the space time where
the curvature is strong (of the order of the Planck scale);
namely in the regions  where the $1/M^2_{string}$ expansion (or
$\alpha'$ expansion) breaks down and the notion of the effective
field theory is ill defined. In the framework of exact string
solutions however, this expansion is not necessary and thus, we can
extend our description at the string level using the powerful
techniques of the underlying two dimensional (super-)conformal
theory. It is then possible (at least in certain cases) to go through
the strong curvature region (where the topology and/or the geometry
change occurs) towards another asymptotic region where we have a
different low
energy field theory. In the exact string solution we have studied,
this  change is described in terms of a modulus field that varies
with time \cite{kk11,tsey} and thus the effective field theory
transitions occur dynamically  during the cosmological evolution.

Here, we have for the first time
the possibility to address an interesting phenomenon such as
dynamical topology change in the context of an exact solution to
classical string theory.

\section{String states and their corresponding effective field
theories}
\setcounter{equation}{0}

We will start from the simple case of a compactification on a flat
torus to illustrate  some typical stringy phenomena and we will
eventually move to discuss  non-flat backgrounds.
The spectrum of string physical states in a given compactification
can be generically separated in three kinds of sub-spectra.

 {\bf i})The first one consists of Kaluza-Klein-like effective field
theory modes (or momentum modes).
The masses of such modes are always proportional to the typical
compactification scale $M_{c}=M_{string}/R$ where $R$ is a
typical radius in $M_{string}$ units (in more general cases, when
there are more than one compact dimensions, one can still
define the concept of a scale \cite{si}).

{\bf ii})The second set of states are the  winding modes which exist
here
because of two reasons. The first is that the string is an extended
object. The second is that the target space has a non-trivial
$\pi_{1}$ so
the string can wind around in a topologically non-contractible way.
In a large volume compact space, these modes are always super-heavy,
since their mass is proportional to the compactification
volume $1/M_{c}$, (more precisely, it is proportional to
$M_{string}^2/
M_{c}$).

{\bf iii})The third class of states are purely stringy states
constructed from the string oscillator operators. Their masses are
always proportional to the string scale $M_{string}=1/\sqrt{\alpha
'}$.

This separation of the spectrum is strictly correct in torroidal
backgrounds. However, further analysis indicates that one can extend
the notions of Kaluza-Klein (KK) type modes and winding modes to at
least non-flat backgrounds with some Killing symmetries.
This generalization comes with the help of the duality symmetries
present in such background fields \cite{bu}-\cite{AG,gk}.
However, the ``winding" states are not always associated with
non-contractible circles of the manifold. They appear as winding
configurations in a (usually) contractible
circle associated with a Killing coordinate \cite{k2}.

Once we have the picture above concerning different types of string
excitations we can state that the notion of a string effective field
theory makes sense  when the ``winding" states as well as the
oscillator states are much heavier than the field theory-like KK
states.
Here, we will assume the presence of a single scale (except
$M_{string}$).
If there are more such scales then one has to investigate the
different regimes. In any such regime, our discussion below is
applicable.

Using the $1/M_{string}^2$-expansion one finds the effective theory
which is
relevant for the low energy processes ($E_t<M_{string}$) among the
massless states, as well as the lowest lying  KK states, provided,
that the typical mass scale $M_c$, (which could be a compactification
scale, gravitational curvature scale etc ) is much below the string
scale $M_{string}$. Otherwise the notion of the effective theory is
not
applicable.

In the last few years, a lot of activity was devoted in understanding
the effective theories of strings at genus zero, \cite{g0} and in
some cases the genus one corrections were included \cite{g1}. The
output of this study confirms that the winding and the ${\cal
O}(M_{string})$ string  superheavy modes can be integrated out and
one
can  define consistently  the string  low energy theory in terms of
the massless and lowest massive KK states. Thus, the perturbative
string solutions are well  described by classical gravity coupled to
some gauge and matter fields with unified gauge, Yukawa and self
interactions. As long as  we stay in the regime where $E_{t},M_c
<M_{string}$ nothing dangerous is happening and consequently our
description of the physics in terms the the effective theory is good
with well defined and calculable ${\cal O}(E_t/M_{string}),{\cal
O}(M_c/M_{string})$ corrections.

The situation above  changes drastically once the mass of  ``winding"
type  states becomes smaller than the string scale and when (usually
at the same time) the KK modes have masses above the string scale.
This is the case when the typical $M_c$ scale is larger than
$M_{string}$. When this occurs, the relevant modes are not any more
the
KK ones but rather the winding ones. Thanks to the well known by now
generalized string duality \cite{bu,k1,GR,k2,AG,gk} (e.g. the
generalization of the well known R to 1/R torroidal duality) it is
possible to find an alternative effective field theory description
by means  of ${\cal O}((E_{t},\tilde M_c )/M_{string})$ dual
expansion, where one uses the dual background which is characterized
by $\tilde M_c$ instead of the initial one characterized by a high
mass scale. The dual mass scale  $\tilde M_c=M^{2}_{string}/M_c$
is small when the $M_c$ becomes big and vice-versa. We observe that
in both extreme cases, either
(i) $M_c < M_{string}< \tilde M_c$ or
(ii) $M_c>M_{string}>\tilde M_c$,  a field theory description exists
in
terms of the original curved background metric in the first case or
in terms of its dual in the second case. This observation is of main
importance since it extends the notion of the effective field
theories in backgrounds with associated high mass scales (due to size
or curvature).

Strictly speaking, there can be   many dual backgrounds  which
correspond to the same string
solution and it is an open problem if it is always possible to map
all regions of spacetime with high associated scales to ones with
small such scales.
In fact, this is not always necessary, since, even for regions which
are strongly curved or have small volume, we can have a well defined
effective description.
Examples could be the region close to special symmetry points in
torroidal compactifications, where although the torus has volume of
order one, we can easily handle the low energy spontaneously broken
gauge theory.
Then, a general string solution would give rise to a $set$ of
$effective$ $field$
$theories$ defined in restricted regions of space-time $(x^{\mu})_I$,
I=1,2,3... with $M_I$  smaller or of the same order as $M_{string}$;
If
$T_{I,J}$ is the boundary region among $(x^{\mu})_I$ and
$(x^{\mu})_J$, then on $T_{I,J}$ we have almost degenerate effective
characteristic scales $M_I\sim M_{string}\sim M_J$.
In such regions, the effective field theory description of regions
$I,J$ break
down (individually), and the full string theory is needed in order to
have a smooth transition between the two.

A goal in that direction would be to establish some simple rules that
would provide the extra stringy
information that would $glue$ the field theoretic regions together.
This effectively amounts to a reorganization of the $\alpha'$
expansion. In simple backgrounds, like torroidal ones this
$gluing$ can be effectively done \cite{gp} by constructing an
effective action, containing an infinite number of fields.

The models we have studied\cite{kk11} could very well
serve as a laboratory towards answering this question.
This is certainly important, since many interesting phenomena happen
precisely at such regions in string theory. We can mention, the
$gluing$ of
$dual$ $solutions$ in cosmological contexts, \cite{bv} and global
effective theories for large regions of internal moduli spaces.
A related issue here is, that with each region, one has an associated
geometry
and spatial topology, as dictated by the effective field theory.
It turns out that moving from one region to another not only the
geometry can change but also the topology. Examples in the context of
Calabi-Yau compactifications
\cite{tc} and more simple models, \cite{gk} have been given.

There is another important point about topology change that we would
like
to stress here. Where topology change happens, depends crucially on
the values of some of the parameters in the background. One such
parameter is always $M_c/M_{string}$,
but usually, in string backgrounds, there are others, like various
different levels for non-simple WZW and their descendant conformal
field theories,
various radii or related moduli etc.
The absolute judge concerning topology change is the effective field
theory.

Another application of such solutions could be in considering strings
at finite temperature.
They would describe a string ensemble with temperature that varies
(adiabatically) in space. It might be interesting to entertain such
an idea in more detail, in order to investigate temperature gradients
in string theory.

\section{Stringy  dynamical transitions among effective field
theories}
\setcounter{equation}{0}

We will present now  a class of exact  string cosmological models
\cite{kk11}
 where the geometry and/or topology of their effective field theories
changes dynamically with the time.
As a starting point we will consider first a static four dimensional
model based
on the following space-time metric:

\begin{equation}
ds^2=-dt^2+k[dx^2+\frac{sin^2x}{cos^2x+R^2sin^2x}d\theta^2_1+
\frac{R^2cos^2x}{cos^2x+R^2sin^2x}d\theta^2_2]
\label{metric} \end{equation}

The above metric corresponds to an exact string solution for any
value of the parameter $R$. For the special value R=1 the space is a
three dimensional sphere ($S^3$) with constant scalar curvature $\hat
R=6/k$. For arbitrary $R~\epsilon~(0,\infty)$ the above metric, from
the conformal theory point of view, corresponds to an exact
deformation (by ~$J_3~{\bar J}_3$) of the $SU(2)_k$
WZW-model with a scalar curvature ${\hat R}$ given by,
\begin{equation}
{\hat R}=-{2\over k}{[2(R^{4}-1)\sin^2 x-5(R^2-1)-3 ] \over
 (1+(R^2-1)\sin^2 x)^2} \; .
\label{r}
\end{equation}
The manifold is regular except at the end-points where
\begin{equation}
{\hat R}(R=0)=-{4\over k\cos^2 x}\;\;,\;\;{\hat R}(R=\infty)=-{4\over
 k\sin^2 x}\; .
\end{equation}
We have to stress here, that even at the end points the string theory
is well defined and thus the curvature singularity of the effective
field theory is an artifact. Indeed, at these points the $SU(2)_k$
deformed theory factorizes to the  $(SU(2)/U(1))_k$ parafermionic
theory and to a $U(1)$ non-compact dimension. Both conformal
sub-systems ($SU(2)_k$ and $U(1)$), have well defined and
non-singular correlations functions although their effective field
theory metric has curvature singularities.
It should be noted also that, the geometric data (metric, curvature,
etc.)
are invariant under the duality transformation $R\rightarrow 1/R$ and
$x\rightarrow \pi/2-x$.

In order to give a more complete description of the above model we
must specify the other two non-trivial background fields; The dilaton
$\phi(x)$ and the antisymmetric tensor field $B_{\mu,\nu}(x)$:
\begin{equation}
\phi(x,R)=log[\frac{cos^2x+R^2~sin^2x}{R^2}]+ \phi_0~~,
\end{equation}
\begin{equation}
B_{\theta_1,\theta_2}(x,R)=\frac{cos^2x}{(cos^2x+R^2~sin^2x)}
\end{equation}

The string energy spectrum (in $M_{string }$ units ) of the model is
given in terms of the deformed $SU(2)_k$ theory quantum numbers; the
$SU(2)$ spin $j$ and  the left and right $j_3,{\bar j}_3$ charges $m$
and $\bar m$

\begin{equation}
E^2(R)=2{j(j+1)\over k+2}-{m^2+\bar m^2\over k}
+{1\over
2k}\left[(m-\bar m+kM)^2
R^2+{(m+\bar m+kN)^2\over R^2}\right]
+N'+\bar N'\label{spec}
\end{equation}
where $N', \bar N'$ denote the string oscillator contributions.

For $k$ relatively large (semi-classical limit) the would be excited
states below the string scale (states with $E^2(R)\ll M^2_{string}$)
are those which correspond to $N'=\bar N'=0 $ and :\\
\\
i) $~~m-\bar m=M=0$ and $(kN)^2<R^2$ if $~R>>k$, \\
ii) $~N=M=0$ and $m, \bar m <<k$ if $~R \sim 1$.\\
iii) $m+\bar m=N=0$ and $(kM)^2< R^{-2}$ if $~R^{-1}>>k$,\\

In these three $R$ intervals the lower lying states are organized in
a different  manner; i) ${\cal O}(1/k)$ and ${\cal O}(k/R)$ expansion
when $R$ is large, ii) ${\cal O}(1/k)$ when $R \sim 1$ and iii)
${\cal O}(1/k)$ and ${\cal O}(kR)$) when $R$ is small:

\begin{equation}
E^2(R>>k)=\frac{2}{k}[j(j+1)-m^2+(m+\frac{kM}{2})^2\frac{1}{R^2}]
\label{speci} \end{equation}

\begin{equation}
E^2(R\sim 1)=\frac{1}{k}[2j(j+1)-m^2-{\bar m}^2+(m-{\bar
m})^2R^2+(m+\bar m)^2\frac{1}{R^2}]
\label{specii}\end{equation}

\begin{equation}
E^2(R<<k)=\frac{2}{k}[j(j+1)-m^2+(m+\frac{kN}{2})^2 R^2]
\label{speciii}\end{equation}

The $R>>1$ effective field theory spectrum is well described by the
$[SU(2)/U(1)]_k$ semi-classical spectrum ($\frac{1}{k}[2j(j+1)-m^2]$)
and by a large radius, almost decompactified, $U(1)$ spectrum
($l^2/R^2$); The effective field theory metric in this regime is
defined by (with $ {\tilde \theta}_1=\theta_1/R$):

\begin{equation}
ds^2(R>>1)\sim -dt^2+k[dx^2+d{\tilde \theta}_1^2+ {cos^2x \over
sin^2x} d\theta_2^2 ]
\label{metrici}\end{equation}

Around $R\sim 1$ the semiclassical spectrum  is similar to a small
deformed $S^3$ sphere and the effective field theory metric is
approximately given by:

\begin{equation}
ds^2(R\sim 1)\sim -dt^2+k[dx^2+sin^2x d \theta_1^2+ cos^2x
d\theta_2^2 ]
\label{metricii}\end{equation}

Finally when $R<<1$ the spectrum is well described by the $dual$
$[SU(2)/U(1)]_k$ semi-classical spectrum ($\frac{1}{k}[2j(j+1)-m^2]$)
and by a dual $U(1)$ spectrum ($l^2~R^2$). The effective theory
metric becomes now: (with ${\tilde \theta}_2=\theta_2~R$)

\begin{equation}
ds^2(R<<1)\sim -dt^2+k[dx^2+{sin^2x \over cos^2x } d\theta_1^2+ d
{\tilde \theta}_2^2 ]
\label{metriciii}\end{equation}

In the above described example the case $R>>1$ and $R<<1$ give
isomorphic effective field theories; they have the same topology and
their metrics are diffeomorfic. This property it is not generic for
other string examples. For
instance, the string example which is based on a deformed $H^+_3$
hyperboloid, instead of the deformed $SU(2)_k$ described above,
defines in $R>>1$ regime an effective field theory which
topologically is different from the effective field theory in the
$R<<1$ regime and that which is defined in $R\sim 1$ region. All
geometrical data on $H^+_3$ can be obtain formally from those of
$SU(2)_k$ by means of the following analytic continuation:
\begin{equation}
t\rightarrow t,~~x\rightarrow ix,~~\theta_1\rightarrow
\theta_1,~~\theta_2 \rightarrow i\theta_2;~~k\rightarrow -k,~~
sinx\rightarrow isinhx,~~cosx\rightarrow coshx.
\end{equation}

The effective theory  metric of the deformed $H^+_3$ model is regular
for $R<<1$ and is factorized in a two sub-spaces. The one sub-space
is that which is described by the semi-classical metric of the
``axial" gauged WZW model $[SU(2)/U(1)]_k$ and has the shape of a
``two-dimensional cigar" ($x,\theta_1$), while the second sub-space
is  described by a non-compact coordinate (${\tilde \theta}_2$). The
effective theory metric around $R\sim 1$ is that of a static
pseudosphere ($H^+_3$ hyperboloide) with constant negative curvature
$\hat R =-6/k$. For $R>1$ the effective theory metric has
singularities at $sinh^2x=1/(R^2-1)$ and for very large $R$ is
factorized and is well described
by the ``vector" gauged WZW model $[SU(2)/U(1)]_k$ together with an
extra  non-compact $U(1)$ factor. The three dimensional space in this
limit is factorized to a two dimensional space with the shape of a
``trumpet" ($x,\theta_2 $) and to the one which is defined by  one
non-compact coordinate ${\tilde \theta}_1$.

We would like to make the static picture described above dynamical
giving to the parameter $R$ an evolution in time. If that can be done
in a stringy way then it will be possible to pass through different
effective field theories during the cosmological evolution. This
dynamical changing phenomenon cannot be described in terms of a
unique effective field theory with finite number of states. We have
shown\cite{kk11}  that,  in both $SU(2)$ and $H^+_3$ case, there
exists a stringy extension with $R$ now a function of time,  which is
adiabatic at least is some time intervals provided that $R(t)$
satisfies the following non linear differential equation.

\begin{equation}
{R''' \over R''}={R'' \over R'}+{R' \over R}
\label{Rdyn}\end{equation}

All solutions of this equation correspond to exact string solutions
and can be classified in terms of two parameters $C_1$ and $C_2$. One
has after integration:
\begin{equation}
{R' \over R}= C_1~R +C_2{1 \over R}
\label{Rdyn2}\end{equation}
The metric and antisymmetric tensor of the static case have the same
form in terms of $R(t)$. The dilaton  field however gets some
corrections:
\begin{equation}
\phi(x,R(t))=log[\frac{cos^2x+R(t)^2~sin^2x}{R(t)^2}]-log[{R'(t)
\over R(t)}] \phi_0~~,
\label{deltaphi}\end{equation}

The solutions with non-trivial $R(t)$ are (up to shifts in $t$):

({\bf ia})  $C_{1}=0$:
\begin{equation}
R^2(t)=C_{2}^{2}t^2\label{ia}
\end{equation}
For $t\in [0,\infty)$, $R^2\in [0,\infty)$.

({\bf ib})  $C_{2}=0$:
\begin{equation}
R^2(t)={1\over C_{1}^2t^2}\label{ib}
\end{equation}
For $t\in [0,\infty)$, $R^2\in [0,\infty)$.

({\bf ii}) $C_{1}C_{2}>0$:
\begin{equation}
R(t)=\sqrt{C_{2}\over C_{1}}\tan(\sqrt{C_{1}C_{2}}t)\label{ii}
\end{equation}
For $t\in [0,\pi/2\sqrt{C_{1}C_{2}}]$,  $R^2\in [0,\infty)$.

({\bf iii}) $C_{1}C_{2}<0$:
\begin{equation}
R(t)=\sqrt{-{C_{2}\over
C_{1}}}\tanh(\sqrt{|C_{1}C_{2}|}t)\;\;\;\;\;{\rm
and}\;\;\;\;\;R(t)=\sqrt{-{C_{2}\over
C_{1}}}\coth(\sqrt{|C_{1}C_{2}|}t)\label{iii}
\end{equation}
Here, for the $tanh$ solution, $R^2\in [0,1]$ whereas for the $coth$
solution
$R^2\in [1, \infty )$.

All of the above solutions correspond to exact conformal field
theories with central charge
$c=4-6\left(1/(k+2)-4C_{1}C_{2}/(3-4C_{1}C_{2})\right)$. When $C_i$
are such that, $c=4$, then their  supersymmetric extension are
nothing but deformations and analytic continuations   of the exact
$N=4$, $\hat c$=4 supercorfomal systems\cite{koun} and which were
used\cite{di}  to construct exact supersymmetric string solutions in
non trivial space-time. The euclidean version of them corresponds to
a class of gravitational and axionic instantons\cite{kkl}.

In summary, exact string solutions as those described above, have a
twofold
interest. First they describe cosmological solutions in which the
effective field theory describing the early stages of our universe is
completely different from the one describing its later stages.
The study of such solutions can provide us with some useful hints for
understanding better the string unification at high scales and  with
quantitative predictions at low energies.

{\bf  Acknowledgements}

We would like to thank I. Bakas, R. Brustein, S. Ferrara, G.
Veneziano, E. Verlinde and F. Zwirner for fruitful discussions.
The work of C. Kounnas was
partially supported by EEC grants SC1$^{*}$-0394C and
SC1$^{*}$-CT92-0789.

\end{document}